\documentclass[sigconf]{acmart} 

\AtBeginDocument{%
  }

\usepackage{multirow}
\usepackage{cuted}
\usepackage{threeparttable}
\usepackage{enumitem}
\usepackage{makecell}

\usepackage[table]{xcolor}
\definecolor{gainsboro}{rgb}{0.86, 0.86, 0.86}
\definecolor{darkviolet}{HTML}{9400D3}

\usepackage{tikz}
\usetikzlibrary{positioning, arrows.meta}

\usepackage[most]{tcolorbox}
\usepackage[noabbrev,capitalize,nameinlink]{cleveref}

\usepackage{framed}
\usepackage{xltabular} 

\usepackage{pifont}
\newcommand{\cmark}{{\color{green!70!black}\ding{51}}}
\newcommand{\xmark}{{\color{red!70!black}\ding{55}}}

\newcommand*\circled[1]{\tikz[baseline=(char.base)]{\node[shape=circle,draw, solid,inner sep=0.5pt] (char) {#1};}}

\newcommand{\pquote}[2]{``\textit{#1}'' -- [#2]}

\newcommand{\simplequote}[1]{``\textit{#1}''}
\newcommand{\cquote}[1]{``\textit{#1}''}

\newlength\bubblesize
\newcommand\yes{\tikz[baseline=0.1ex] \fill[black]
(\bubblesize,\bubblesize) circle (\bubblesize);}
\newcommand\may{\begin{tikzpicture}[baseline=0.1ex, line width=0.2ex]
    \draw[clip]  (\bubblesize,\bubblesize) circle (\bubblesize-.5\pgflinewidth);
    \fill[black] (\bubblesize, 2\bubblesize) rectangle (0,0);
  \end{tikzpicture}
}
\newcommand\no{\tikz[baseline=0.1ex] \draw[black, line width=0.2ex] (\bubblesize,\bubblesize) circle (\bubblesize-.5\pgflinewidth);}

\makeatletter
 
\renewcommand*\textcircled[1]{
  \tikz[baseline=(char.base)]{
    \node[shape=circle,draw,inner sep=0.5pt] (char) {#1};}}
\makeatother

\newcounter{solution}
\newenvironment{solution}{
  \refstepcounter{solution}\par\medskip
  \begin{tcolorbox}[
    colback=gray!10,
    colframe=black, 
    arc=4pt,
    boxrule=0.5pt,
    left=6pt, right=6pt, top=6pt, bottom=6pt
  ]
  \noindent\textbf{Main Takeaways:} \rmfamily
}{
  \end{tcolorbox}
  \medskip
}

\renewcommand\footnotetextcopyrightpermission[1]{}

\begin{document}

\title{``They don't care about this'': A Systematic Study of TEE Build Reproducibility in the Wild}

\author{Annika Wilde}
\affiliation{%
  \institution{Ruhr University Bochum}
  \city{Bochum}
  \country{Germany}}
\email{annika.wilde@rub.de}

\author{Marco Gutfleisch}
\affiliation{%
  \institution{LMU Munich}
  \city{Munich}
  \country{Germany}}
\email{marco.gutfleisch@ifi.lmu.de}

\author{Felix Reichmann}
\affiliation{%
  \institution{Ruhr University Bochum}
  \city{Bochum}
  \country{Germany}}
\email{felix.reichmann@rub.de}

\author{Anirban Chakraborty}
\affiliation{%
  \institution{Max Planck Institute for Security and Privacy}
  \city{Bochum}
  \country{Germany}}
\email{anirban.chakraborty@mpi-sp.org}

\author{Yuval Yarom}
\affiliation{%
  \institution{Ruhr University Bochum}
  \city{Bochum}
  \country{Germany}}
\email{yuval.yarom@rub.de}

\author{M. Angela Sasse}
\affiliation{%
  \institution{Ruhr University Bochum}
  \city{Bochum}
  \country{Germany}}
\email{martina.sasse@rub.de}

\author{Ghassan Karame}
\affiliation{%
  \institution{Ruhr University Bochum}
  \city{Bochum}
  \country{Germany}}
\email{ghassan@karame.org}

\begin{abstract}
    Trusted Execution Environments (TEEs) have become a cornerstone of modern cloud computing, providing strong confidentiality and integrity guarantees for both code and data. 
    A critical component of this trust model is remote attestation, which enables external entities to verify the authenticity and integrity of code executing within a TEE through cryptographic measurements. 
    However, the effectiveness of remote attestation fundamentally depends on the verifier's ability to trace the reported measurement back to the original source code---a property that can only be guaranteed through reproducible builds.
    
    In this paper, we investigate the reproducibility of TEE builds through a technical analysis of 115 TEE deployments. Our analysis spans popular TEEs such as Intel SGX, Intel TDX, and AMD SEV, and reveals that a striking 91\% of those deployments were not reproducible, with 80\% failing to provide both source code and a reference build, the two essential prerequisites for reproducibility. 
    To explore the root causes, we contacted the maintainers of 50 SGX projects and managed to recruit 12 developers from industry and academia for interviews.  Only one of our participants reported that reproducibility is a priority during development, effectively confirming our technical findings. Beyond technical barriers (e.g., timestamps included in the binary) that can be readily addressed, we identify broader ecosystem-level challenges, such as the lack of control over the build environment in projects involving multiple stakeholders. 
    We argue that achieving reproducibility in TEEs requires a holistic development approach that extends beyond individual developers and calls for stronger commitments---rather than treating TEEs as a ``security badge''.
\end{abstract}

\maketitle

\pagestyle{plain}

\section{Introduction}

The rapid adoption of cloud computing has increased the demand for distributed and decentralized applications, where infrastructure resources---including hardware, platforms, and software---are owned and operated by different entities. This paradigm shift in the IT landscape has also intensified concerns about the confidentiality and integrity of sensitive data and computations performed on remote systems.
Confidential computing has emerged as a promising approach to address these challenges and ensure data protection in the cloud. At its core, confidential computing relies on \emph{Trusted Execution Environments (TEEs)}, specialized hardware components within modern CPUs that create isolated regions of execution, often called \emph{enclaves} or \emph{confidential virtual machines (CVMs)}. These environments safeguard both the data being processed and the integrity of the code performing the computation, offering strong assurances even when the surrounding system or cloud provider cannot be fully trusted.

To provide these security guarantees, TEEs minimize the Trusted Computing Base (TCB)---the set of components that must be trusted to remain secure---to contain only the CPU, the TEE firmware, and the code running inside the enclave. 
Users can verify the integrity of this TCB and the code running inside the enclave through \emph{remote attestation} (details in \cref{sec:bg_tees}). %
As such, they are expected to entrust their sensitive data to an enclave if and only if its attested measurement matches a known, trusted reference.

The TEE ecosystem has enabled a wide range of applications, including secure messaging platforms, such as Signal~\cite{signal}, privacy-preserving databases~\cite{mongodbsgx_repo}, distributed machine learning~\cite{blindai}, and blockchain systems~\cite{secret_network_website,ternoa_website}, that leverage the ability to delegate computation to remote infrastructure while maintaining strong guarantees of data confidentiality and integrity.
For example, consider a privacy-preserving machine learning (PPML) service~\cite{secretflow} that processes anonymized medical data using open-source models hosted on a remote server. Multiple users, such as hospitals, diagnostic centers, insurance companies, etc., can securely contribute their datasets to train a shared model for tasks like early disease detection or patient health assessment. The security assurances provided by TEEs ensure that data is protected, even from strong adversaries, such as compromised or corrupt cloud providers.

Although remote attestation guarantees that the application running in the enclave is signed and authenticated by the developer, it does not provide any assurances about the trustworthiness of the application code itself. A key challenge in the adoption of TEEs is ensuring that the code executed inside the enclave is trustworthy and free from malicious backdoors or vulnerabilities that could leak sensitive data to the application developer. 
In essence, TEEs aim to shift trust from developers to code: rather than relying on developers' integrity, users can inspect and verify the enclave code to confirm that it meets required security and privacy guarantees.

Returning to the example of the PPML service, before sharing private patient information, users could examine the open-source enclave code to ensure it does not intentionally or inadvertently expose sensitive data. 
Similarly, for other sensitive applications, such as cryptocurrencies, smart contracts, or electronic-voting systems, it is essential to validate the trustworthiness of the code before entrusting the service with sensitive information.

\begin{figure}[tbp]
    \centering
    \includegraphics[scale=0.52]{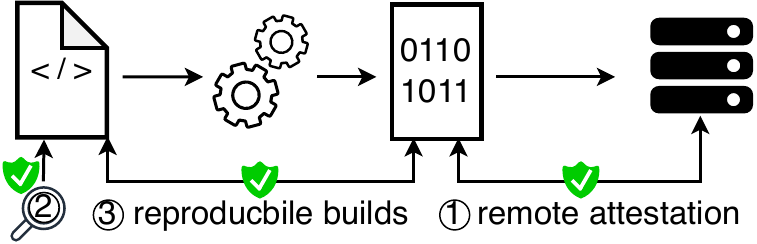}
    \caption{Interplay between remote attestation and reproducible builds. Remote attestation links a service to the specific deployed binary (Step~\protect\circled{1}), while source code inspection (Step~\protect\circled{2}) and reproducible builds (Step~\protect\circled{3}) establish trust in the source code and connect it to the attested binary.}
    \label{fig:ml_sample}
\end{figure}

Once users are satisfied with the source code, they must be able to confirm that the deployed binary was genuinely built from the vetted source code. This requires \textbf{linking the enclave's measurement to the source code}, typically by recompiling the source locally and comparing the resulting binary to the attested one.\footnote{Note that the process of auditing source code and compiling it to derive a trusted reference measurement is inherently time-consuming and requires technical expertise that exceeds the capabilities of most end users. However, such verification need not be performed by every individual. A subset of technically skilled and security-conscious users can conduct these steps and publish verified reference measurements for the community. In the PPML example, this collaborative verification model allows ordinary users to rely on community-vetted attestations, establishing collective trust in the service's confidentiality guarantees.} For this comparison to succeed, the locally compiled binary must be \emph{bit-by-bit identical} to the deployed version. Achieving such identity depends on \emph{reproducible builds}---ensuring that identical inputs and environments always yield identical outputs. \Cref{fig:ml_sample} shows such a verification process, depicting the interplay between remote attestation and reproducible builds.

The example above highlights that the reproducibility of the build process is a critical component of TEE security. However, it remains unclear whether state-of-the-art and widely deployed TEEs actually support reproducible builds. Furthermore, there is limited understanding of whether TEE developers themselves recognize the importance of reproducibility in ensuring software integrity and trust.

In this paper, we present the first comprehensive study of reproducibility in TEE builds, combining two complementary approaches: \textcircled{1} an empirical analysis of 115 TEE projects drawn from~\cite{awesome_sgx,azure,gcp}, covering widely used TEEs including Intel SGX, Intel TDX, and AMD SEV-SNP; and \textcircled{2} a qualitative interview study with 12 SGX developers. Our findings paint a stark picture: reproducible builds remain the exception rather than the norm across today's TEE ecosystem. Of the 115 projects examined, 94\% lack the documentation or tooling needed to reproduce enclave measurements, and 77\% fail to provide both source code and a reference build---two fundamental prerequisites for reproducibility. Even when developers supplied additional information directly, 91\% of the projects remained unreproducible. We further demonstrate that these challenges are not specific to any particular TEE technology, persisting across SGX, TDX, and SEV-SNP alike; however, we observe that reproducibility rates vary depending on the primary build system used, with notable differences observed across systems such as prebuilt Docker images and dependency-pinning tools such as Nix~\cite{nix}  (cf.\ \cref{sec:reproducibility_in_the_wild}). These results are echoed in our interview study, where 10 of the 12 developers reported that reproducibility is neither prioritized nor considered during development (cf.\ \cref{sec:interview_study}).

We analyze the reasons behind this lack of reproducibility and identify several issues that prevent users from reproducing the enclave binary, as well as challenges faced by developers. To our surprise, many of the technical obstacles, such as embedded time\-stamps or non-deterministic file system ordering, seem to be well-known in the broader, traditional reproducible-builds community. Other reasons include an occasional inclusion of secret material within the enclaves, which cannot be reproduced by anyone outside the project.
Beyond these technical factors, our qualitative study uncovers several ecosystem-level and procedural challenges. In particular, the traditional software development model process, where components are built in isolation, is incongruent with the requirements of TEE development. Since every component in the build pipeline can influence the enclave's final measurement, developers must adopt a broader, system-level view of the build process. This mismatch undermines one of the core security mechanisms of TEEs: if an enclave's measurement cannot be reproduced, remote attestation loses its true meaning (cf.\ \cref{sec:discussion}).
We conclude that ensuring build reproducibility in TEE applications requires more than simple technical solutions---it demands a rethinking of development practices. We therefore argue for a TEE-specific development process that explicitly prioritizes reproducibility and enables developers to regain control over their build environments to ultimately seize the value of cloud computing  (cf.\ \cref{sec:recommendations}).

\section{Background and Related Work}

\subsection{Trusted Execution Environments}
\label{sec:bg_tees}

Trusted Execution Environments (TEEs) use hardware-based mechanisms to protect memory regions at runtime, enabling secure execution of sensitive code in isolated environments. Depending on their granularity, TEEs can isolate either a single process---known as \emph{process-based TEEs}~\cite{DBLP:conf/isca/HoekstraLPPC13/sgx, whitepaper:trustzone, eurosys/LeeKSAS20/keystone}---or an entire virtual machine (VM)---referred to as \emph{VM-based TEEs}~\cite{whitepaper:sevSnp, whitepaper:tdx}. Examples include Intel Software Guard Extensions (SGX)~\cite{DBLP:conf/isca/HoekstraLPPC13/sgx} and Arm TrustZone~\cite{whitepaper:trustzone} for process-based TEEs, and Intel Trust Domain Extensions (TDX)~\cite{whitepaper:tdx} and AMD Secure Encrypted Virtualization (SEV) Secure Nested Paging (SNP)~\cite{whitepaper:sevSnp} for VM-based TEEs. In this work, we generally refer to a TEE instance initialized with user code as an \emph{enclave}.

The TEE trust model assumes that only the CPU, TEE firmware, and the enclave's specific code form the Trusted Computing Base (TCB). In other words, all privileged system software, such as the operating system and hypervisor, is excluded from the TCB and treated as untrusted, since it could potentially be compromised. 

Each enclave is characterized by two cryptographic identifiers. The first, referred to as \emph{measurement}, is a hash of the enclave's initial code and data, uniquely defining its content. The second, called \emph{signer identity} in Intel TEEs, is derived from the developer's public key and uniquely identifies the entity that signed the enclave.
To verify an enclave's authenticity, TEEs such as Intel SGX, Intel TDX, and AMD SEV provide a \emph{remote attestation} mechanism. During attestation, the TEE firmware generates a signed certificate of the enclave's measurement using a hardware-specific key. A remote verifier can validate the enclave's integrity by checking this signature through an attestation service (e.g., one operated by the hardware vendor) and comparing the reported measurement to a known trusted reference. In contrast to Intel SGX and AMD SEV-SNP, which report a single measurement in the attestation, Intel TDX exposes five measurement registers; four of them can be updated at runtime. Successful verification confirms that the enclave is executing the expected code within a genuine TEE, thereby upholding the TEE's security guarantees.

\subsection{Reproducible Builds}
\label{sec:reproducible_builds}

Reproducible (or deterministic) build is a software development practice that ensures a build process consistently produces identical output across different environments. In other words, when a build is reproducible, the same source code, compiler version, and build configuration always yield the same output, regardless of where or when the build is performed.
Build reproducibility can be categorized into two types: \emph{bit-by-bit reproducibility} and \emph{semantic reproducibility}. Bit-by-bit reproducibility guarantees that two independently produced binaries are identical at the byte level, while semantic reproducibility is less strict, requiring only identical behavior, even if the binary representations differ.

Reproducibility is critical for verifying the integrity and authenticity of software binaries. It allows developers and users to independently rebuild software and confirm that the resulting binary matches the expected output. This is particularly important for open-source projects, where transparency and verifiability are key---anyone can inspect the source code, reproduce the build, and confirm its trustworthiness. By supporting reproducible builds, developers strengthen transparency and trust in their software.

\paragraph{Technical Challenges in Reproducing Builds} Traditional build processes depend on numerous environmental and non-deter-ministic factors, such as the operating system, file system, and even timestamps. These sources of variability introduce non-determinism that must be mitigated to achieve reproducibility. Lamb and Zacchiroli~\cite{DBLP:journals/software/LambZ22/rbTechnicalChallenges} identify several common sources:

\begin{itemize}[nosep,left=0pt]
\item \textbf{Timestamps:} Macros like \texttt{\_\_DATE\_\_} often embed build timestamps into the binary, creating variability between two builds produced at different points in time.
\item \textbf{Build paths:} Macros like \texttt{\_\_FILE\_\_} record absolute file paths in binaries, which vary between build environments unless a fixed directory structure is enforced.
\item \textbf{File system ordering:} File systems may order directory contents differently, causing inconsistent file processing during the build.
\item \textbf{Archive metadata:} Archive %
metadata, such as timestamps and ownership, can differ between builds.
\item \textbf{Compiler randomness:} Some compilers generate random identifiers or values during compilation, introducing non-determinism.
\item \textbf{Uninitialized memory:} In low-level languages (e.g., C/C++), uninitialized memory can cause inconsistent build behavior.
\end{itemize}

\subsection{Related Work}

To the best of our knowledge, no prior work has systematically examined build reproducibility within TEEs. That said, several studies have explored reproducibility in traditional software contexts.
Lamb and Zacchiroli~\cite{DBLP:journals/software/LambZ22/rbTechnicalChallenges} provide a comprehensive overview of sources of non-determinism in build processes and discuss the technical barriers to achieving reproducible builds. 
Similarly, de Carn\'e de Carnavalet and Mannan~\cite{DBLP:conf/acsac/CarnavaletM14/ChallengesOfVerifiableBuilds} analyze 16 versions of the open-source encryption tool TrueCrypt and identify multiple causes of non-determinism that hinder verifiable builds.

Fourné \emph{et al.}~\cite{DBLP:conf/sp/FourneWEFA23/FlossingYourTeeth} investigate the motivations and challenges faced by open-source developers pursuing reproducible builds \linebreak through interviews with 24 contributors of the \emph{Reproducible-Builds.\linebreak org} project. Their findings highlight both technical difficulties, as noted by Lamb and Zacchiroli~\cite{DBLP:journals/software/LambZ22/rbTechnicalChallenges}, and broader ecosystem challenges like limited awareness and poor documentation. 
In contrast, our study focuses on SGX developers to assess their awareness, understanding, and attitudes toward reproducible builds in TEEs. Moreover, we target a community that is inherently interested in reproducible builds, which is essential for the attestation process.

In the context of TEEs, Hugenroth \emph{et al.}~\cite{DBLP:journals/corr/abs-2505-02521/attestableBuilds} propose \textit{attestable builds}, where the build process and input verification occur within a TEE, which then produces a cryptographic attestation of the resulting binary’s hash. 
Similarly, Delignat-Lavaud \emph{et al.}~\cite{DBLP:journals/cacm/DelignatLavaudFVCRCR24/CCFreproducibleBuildArticle} introduce a Code Transparency Service for tracking code provenance and holding developers accountable for released binaries. Both approaches assume developers’ interest in achieving reproducible builds. 
In contrast, our work examines the actual awareness, motivation, and challenges faced by SGX developers, offering insights and recommendations to foster reproducibility within TEE ecosystems.

\section{Reproducible TEE Builds in the Wild}
\label{sec:reproducibility_in_the_wild}

We now evaluate the reproducibility of TEE builds. To this end, we conduct a systematic analysis of TEE-based open-source applications, focusing on their ability to deterministically reproduce enclave measurements.

\subsection{Methodology}
\label{sec:technical_methodology}

\paragraph{Project selection} In our analysis, we selected candidate projects independently from two established sources to minimize selection bias: (i) the marketplaces from Microsoft Azure~\cite{azure} and Google Cloud Platform (GCP)~\cite{gcp}, and (ii) the GitHub repository \emph{Awesome SGX Open Source}~\cite{awesome_sgx} (which, to the best of our knowledge, provides the most comprehensive list of publicly available TEE applications). In the former, we identified candidate projects by searching for the terms \emph{TEE}, \emph{Trusted Execution Environment}, \emph{SGX}, \emph{TDX}, and \emph{SEV-SNP}, discarding products that appeared solely due to loose search policies rather than genuine support or implementation of these technologies. 
Our initial candidate set consisted of 232 applications from~\cite{awesome_sgx} and 35 products from the marketplaces---specifically, seven products from the GCP marketplace~\cite{gcp} and 33 products from the Azure marketplace~\cite{azure}, with five products available on both platforms.
We filtered this candidate set by applying the following three criteria (C1--C3):
\begin{description}[leftmargin=0.4cm]
    \item[C1:] Suitable projects for our study must run within a TEE or contribute to the binary running in the TEE.
    \item[C2:] %
    Marketplace products must be directly purchasable on the platform (i.e., offers requiring prior vendor interaction are excluded). This ensures that relevant technical details are accessible without additional communication. 
    \item[C3:] We are interested in active projects (which we define as having a commit on the default branch after January 2020, not being archived by the end of December 2025, and not being marked as unsuitable for production use).
\end{description}

\begin{table}[t]
    \centering
    \small
    \caption{Overview of dataset filtering criteria (C1--C3). Each row reports the number of projects that met each criterion individually. The last row shows how many met all of them. }%
    \label{tab:selection_results}
    \scalebox{1.0}{
    \begin{tabular}{|l|c|c|}
        \hline
        \textbf{Stage} & \textbf{Awesome SGX} & \textbf{\makecell{Azure \& Google cloud \\ marketplaces}} \\
        \hline
        Initial candidate set & 232 & 35 \\
        TEE-integrated (C1) & 203 & 27 \\
        Directly purchasable (C2) & NA & 29 \\
        Active (C3) & 122 & 35 \\
        Final dataset (C1--C3) & 95 & 21 \\
        \hline
    \end{tabular}
    }
\end{table}

After applying these criteria (cf. \Cref{tab:selection_results}), our final dataset comprises 115 TEE applications: 95 from~\cite{awesome_sgx} and 21 from the marketplaces, with one application present in both sets.
Six applications appear multiple times among the 115 due to multi-TEE support: three support both SGX and SEV-SNP, two support both TDX and SEV-SNP, and one supports all three TEEs.
In our analysis, we treat the same application on different TEEs as separate observations, since reproducibility can vary across deployments, even among projects from the same vendor, as shown by the cases of CCF and the Secret Network (cf. \Cref{sec:secret_network}).
We describe our analysis process below.

\paragraph{Analysis workflow}
To evaluate reproducibility, our analysis is necessarily limited to open-source applications, as access to the source code is required to rebuild the enclave. We therefore consider only applications that provide both source code (with build instructions) and a reference measurement, either as a concrete measurement value or as a release binary. Out of the 115 TEE applications in our dataset, 88 do not meet this criterion and are therefore unreproducible. Conversely, 27 applications fulfill this requirement (cf. \Cref{tab:tee_features}).%

We assess the reproducibility for each of these projects as follows. We first attempted to compile the latest release of each selected project using the build instructions provided in its documentation. If compilation failed, we contacted the corresponding developers for assistance via the official listed communication channels (e.g., email, Discord, or GitHub issues), following this order of preference.

For each successfully compiled application---either independently or with developer support---we analyzed the reproducibility of the resulting enclave build. Specifically, we extracted the enclave measurement from the compiled artifacts using the tools provided by the respective SDK. For the official Intel SGX SDK, for example, we employed the \texttt{sgx\_sign} utility to retrieve enclave metadata, including the measurement. When an explicit reference measurement was unavailable, we derived it from the provided release binary or the official Docker image.

Finally, we compared the measurement of our compiled artifact with the corresponding reference value. When discrepancies occurred, we engaged with developers to identify possible causes of variation and obtain guidance on achieving reproducibility. 
For example, a measurement mismatch occurred when reproducing CCF's SGX enclave binary despite successful compilation. We emailed the repository's contact %
with details of the release, build procedure, and environment, requesting clarification on possible reproducibility issues. Similar messages were sent to other teams---e.g., to Phala~\cite{phala_repo} via email and Ternoa~\cite{ternoa_website} through a Discord \emph{Ticket}.

\subsection{Results}
\label{sec:technical_results}

\Cref{tab:tee_features} provides an overview of the selected projects, while \Cref{tab:study} summarizes our analysis of the 27 TEE applications that provide both source code and a reference measurement. 
We evaluated each application in two scenarios: an \emph{unguided scenario}, where we attempted to build and reproduce the enclave without explicit developer assistance, and a \emph{guided scenario}, in which we engaged with the developers to achieve a successful and/or reproducible enclave build if the unguided attempt failed.
Each entry in~\cref{tab:study} reports (for both scenarios) whether the application was successfully compiled, whether the enclave hash was reproduced, and whether the development team responded to our inquiries. 
We successfully compiled 25 projects independently; the remaining two could not be compiled due to dependencies being unavailable. 

Among the evaluated projects, ten provide sufficient information to reproduce the enclave measurement: seven rely solely on public documentation, while three require non-public information from the developers. %
\emph{In other words, 91.3\% of applications in our dataset are not reproducible. Specifically, 77\% fail to provide both source code and a reference measurement. More surprisingly, among those projects that do provide a reference measurement, only 37\% are reproducible.}

\vspace{0.5em}\noindent\textbf{Academic vs. commercial projects.}
Among the analyzed applications, 27\% are academic artifacts, all but one based on SGX (cf.\ \Cref{tab:tee_features}). 
Reproducibility is 0\% for academic projects and 12\% for non-academic ones. 
Notably, only a single academic project provides a reference measurement, a prerequisite for reproducibility.
This suggests that, as expected, reproducibility is more naturally incentivized in production settings, while academic artifacts are often prototypes.
Given the dataset's imbalance (only 8.7\% reproducible) and skew toward SGX (82.6\%), we restrict our analysis in what follows to the 84 non-academic projects, where reproducibility is more relevant.

\begin{table}[t]
    \centering
    \footnotesize
    \caption{Overview of the selected projects.
    }
    \label{tab:tee_features}
    \scalebox{0.95}{\begin{tabular}{|l|c|c|c|c|}
        \hline
        \textbf{TEE Technology} & \textbf{\rule{0.75em}{0pt}SGX\rule{0.75em}{0pt}} & \textbf{SEV-SNP} & \textbf{\rule{0.75em}{0pt}TDX\rule{0.75em}{0pt}} & \textbf{Unspecified} \\
        \hline
        
        \rowcolor{gray!10}
        \multicolumn{5}{|l|}{\textbf{Dataset}}\\ \hline
        Total projects & 95 & 15 & 3 & 2 \\
        \hline
        \rowcolor{gray!10}
        \multicolumn{5}{|l|}{\textbf{Project type}}\\ \hline
        Commercial & 65 & 14 & 3 & 2 \\
        Academic & 30 & 1 & 0 & 0 \\
        \hline
        \rowcolor{gray!10}
        \multicolumn{5}{|l|}{\textbf{Primary build system}}\\ \hline
        Unspecified & 74 & 8 & 1 & 2 \\
        Docker file & 11 & 3 & 0 & 0 \\
        Docker image & 8 & 1 & 0 & 0 \\
        Yocto & 0 & 1 & 1 & 0 \\
        Bazel & 1 & 1 & 1 & 0 \\
        Nix & 1 & 1 & 0 & 0 \\
        \hline
        \rowcolor{gray!10}
        \multicolumn{5}{|l|}{\textbf{Reference availability}}\\ \hline
        Neither source code nor reference & 14 & 5 & 1 & 2 \\
        Source code only & 57 & 2 & 0 & 0 \\
        Reference only & 2 & 4 & 1 & 0 \\
        \textbf{Source code + reference} & \textbf{22} & \textbf{4} & \textbf{1} & \textbf{0} \\
        \hline
        \rowcolor{gray!10}
        \multicolumn{5}{|l|}{\textbf{Reproducibility}}\\ \hline
        Unreproducible & 89 & 11 & 3 & 2 \\
        Reproducible guided scenario & 3 & 0 & 0 & 0 \\
        Reproducible unguided scenario & 3 & 4 & 0 & 0 \\
        \hline

    \end{tabular}}
\end{table}

\vspace{0.5em}\noindent\textbf{Is there a correlation between the TEE technology (i.e., SGX, TDX, SEV-SNP) and reproducibility?}
Among the filtered projects, Intel SGX leads with 65 commercial implementations, followed by AMD SEV-SNP with 14 and Intel TDX with three. This distribution reflects the relative maturity of each technology: Intel TDX and AMD SEV-SNP are newer additions to the landscape (TDX being the most recent) and are still gaining market traction, a trend that is equally reflected in academic projects as well.

We observe that reproducibility varies notably across these platforms: none of the TDX (most recent TEE) projects are reproducible, only 9.2\% of SGX projects (oldest and most widely studied TEE) are, whereas 28.6\% of SEV-SNP (released somewhere between the two) projects are reproducible.
To assess whether this variation reflects a genuine association, we test the null hypothesis that reproducibility rates are equal across TEE technologies. Fisher's exact test with the Freeman–Halton extension yields a p-value of 0.099, \emph{indicating no statistically significant relationship between TEE technology and reproducibility}.

\vspace{0.5em}\noindent\textbf{Does the choice of the build system correlate with reproducibility?}
We classify non-academic projects according to their primary build system, distinguishing four categories: dedicated reproducibility-oriented tools such as Bazel, Nix, or Yocto (seven projects); prebuilt Docker images (nine projects); Dockerfile-based builds (eleven projects); and unspecified systems (57 projects). The highest reproducibility rate, 42.9\%, belongs to the first category, consistent with these tools' native support for dependency pinning and hermetic build environments. Projects relying on prebuilt Docker images or Dockerfiles achieve lower yet comparable rates of 22.2\% and 18.2\%, respectively, while only 5.3\% of projects with unspecified primary build systems are reproducible.
Fisher's exact test with the Freeman–Halton extension decisively rejects the null hypothesis of equal reproducibility rates across build systems (p = 0.012), \emph{confirming a statistically significant association between build system choice and reproducibility}.

\begin{solution} Our analysis reveals that the challenges of achieving reproducibility (approximately 91\% of all selected TEE projects were not reproducible) are orthogonal to the underlying TEE technology (i.e., SGX, TDX, SEV-SNP). Build system choice, by contrast, emerges as a strong predictor; in fact, dependency-pinning tools such as Nix and Yocto attain the highest reproducibility rate in our dataset (42.9\%).
\end{solution}

\newcommand{\resultRow}[7]{#1 & #2 & #3 & #4 & #5 & #6 \\ }
\begin{table}[!t]
	\caption{Summary of our analysis of 27 TEE applications. The project source is denoted by $^{a}$ (\cite{awesome_sgx}) and $^{m}$ (\cite{azure,gcp}). We report the reproducibility of the enclave hash in an unguided scenario (independently) and a guided scenario (with developer assistance). In the guided case, ``--'' indicates success without assistance; we additionally report whether developers responded. $\star$ indicates inherently unreproducible projects.} %
	\label{tab:study}
	\centering

	\scalebox{0.95}{
    \footnotesize
	\begin{tabular}{ |l|c|c|c|c|c| }
		\hline
		\multirow{2}{*}{\textbf{Project}} &  \multicolumn{2}{c|}{\textbf{Unguided}} & \multicolumn{3}{c|}{\textbf{Guided}} \\
        \cline{2-6}
		  & \textbf{Built}
        & \textbf{Reproduced}
        & \textbf{Replied}
        & \textbf{Built}
        & \textbf{Reproduced} \\
		\hline

        \rowcolor{gray!10}
		\multicolumn{6}{|c|}{\textbf{Intel SGX}} \\ 
        \hline
		\resultRow{CCF \cite{ccf_repo_sgx} \textsuperscript{$m$}}{\cmark}{\xmark}{\cmark}{--}{\cmark}{\cmark}
		\resultRow{Edgeless RT \cite{edgeless_rt_repo} \textsuperscript{$a$}}{\cmark}{\xmark}{\xmark}{--}{\xmark}{\cmark}
		\resultRow{EGo \cite{ego_repo} \textsuperscript{$a$}}{\cmark}{\cmark}{--}{--}{--}{\cmark}
		\resultRow{enclave-vrf \cite{enclave-vrf_repo} \textsuperscript{$a$}}{\cmark}{\xmark}{\xmark}{--}{\xmark}{\cmark}
		\resultRow{Google Asylo \cite{google_asylo_repo} \textsuperscript{$a$}}{\cmark}{\xmark}{\xmark}{--}{\xmark}{\cmark}
		\resultRow{Gramine \cite{gramine_repo} \textsuperscript{$a$}}{\cmark}{\xmark}{\cmark}{--}{\cmark}{\cmark}
		\resultRow{Marblerun \cite{marblerun_repo} \textsuperscript{$a$}}{\cmark}{\xmark}{\cmark}{--}{\cmark}{\cmark}
		\resultRow{MobileCoin \cite{mobilecoin_repo} \textsuperscript{$a$}}{\cmark}{\xmark}{\xmark}{--}{\xmark}{\cmark}
		\resultRow{MongoDB-SGX \cite{mongodbsgx_repo} \textsuperscript{$m$} $\star$}{\cmark}{\xmark}{\cmark}{--}{\xmark}{\xmark}
		\resultRow{Mystikos \cite{mystikos_repo} \textsuperscript{$a$}}{\cmark}{\xmark}{\xmark}{--}{\xmark}{\cmark}
		\resultRow{Nginx-SGX \cite{nginxsgx_repo} \textsuperscript{$a$} $\star$}{\cmark}{\xmark}{\cmark}{--}{\xmark}{\xmark}
		\resultRow{Oasis Sapphire \cite{sapphire_repo} \textsuperscript{$a$}}{\cmark}{\xmark}{\xmark}{--}{\xmark}{\cmark}
		\resultRow{Occulum \cite{occlum_repo} \textsuperscript{$a$}}{\cmark}{\xmark}{\xmark}{--}{\xmark}{\cmark}
		\resultRow{Phala \cite{phala_repo} \textsuperscript{$a$}}{\cmark}{\xmark}{\xmark}{--}{\xmark}{\cmark}
		\resultRow{Safeheron Arweave \cite{safeheron_repo} \textsuperscript{$a$}}{\xmark}{\xmark}{\cmark}{\xmark}{\xmark}{\cmark}
		\resultRow{Linux SGX \cite{linuxsgx_repo} \textsuperscript{$a$}}{\cmark}{\cmark}{--}{--}{--}{\cmark}
		\resultRow{SGXSSE \cite{sgxsse_repo} \textsuperscript{$a$}}{\xmark}{\xmark}{\cmark}{\xmark}{\xmark}{\xmark}
		\resultRow{Secret Network \cite{secret_network_repo_sgx} \textsuperscript{$am$} $\star$}{\cmark}{\xmark}{\cmark}{--}{\xmark}{\cmark}
		\resultRow{Signal CDSI \cite{signal_repo} \textsuperscript{$a$}}{\cmark}{\cmark}{--}{--}{--}{\cmark}
		\resultRow{SnowHaze zka-sgx \cite{snowhaze_repo} \textsuperscript{$a$}}{\cmark}{\xmark}{\xmark}{--}{\xmark}{\cmark}
		\resultRow{Teaclave SDK \cite{teaclave_sdk} \textsuperscript{$a$}}{\cmark}{\xmark}{\xmark}{--}{\xmark}{\cmark}
		\resultRow{Ternoa enclaves \cite{ternoa_enclaves_repo} \textsuperscript{$a$}}{\cmark}{\xmark}{\cmark}{--}{\xmark}{\cmark}

        \hline 
        \rowcolor{gray!10}
		\multicolumn{6}{|c|}{\textbf{AMD SEV-SNP}} \\ 
        \hline
		\resultRow{CCF \cite{ccf_repo_cvm} \textsuperscript{$a$}}{\cmark}{\cmark}{--}{--}{--}{\cmark}
		\resultRow{Cosmian KMS \cite{cosmian_kms_repo} \textsuperscript{$m$}}{\cmark}{\cmark}{--}{--}{--}{\xmark}
		\resultRow{Secret Network \cite{secret_network_repo_cvm} \textsuperscript{$a$}}{\cmark}{\cmark}{--}{--}{--}{\cmark}
		\resultRow{TamaGo \cite{tamago-go_repo} \textsuperscript{$a$}}{\cmark}{\cmark}{--}{--}{--}{\xmark}
        
        \hline 
        \rowcolor{gray!10}
		\multicolumn{6}{|c|}{\textbf{Intel TDX}} \\ 
        \hline
		\resultRow{Secret Network \cite{secret_network_repo_cvm} \textsuperscript{$a$}}{\cmark}{\xmark}{\xmark}{--}{\xmark}{\cmark}

		\hline

	\end{tabular}}
\end{table}

\subsection{Case Studies}
\label{subsec:pitfalls}

During our analysis, we identified 20 unreproducible projects in the unguided scenario. We contacted the developers of the corresponding projects and recovered the root cause in nine cases. We discuss the identified reproducibility blockers in selected case studies below and summarize them in \Cref{tab:root_causes}.

\paragraph{Secret Network}
\label{sec:secret_network}
Secret Network~\cite{secret_network_repo_sgx} is a blockchain platform that leverages TEEs for confidential smart contract execution, using Intel SGX to protect both execution integrity and contract state.
The network promises that contracts are ``private-by-default'' and ``can't be viewed by others''~\cite{secret_network_website}.
The Secret Network node is offered as an Azure application for one-click deployment~\cite{secret_network_azure} and its source code is publicly available, allowing users to inspect it, e.g., to ensure the absence of vulnerabilities or backdoors that may allow the developers to leak secret information. 

However, the Secret Network source repository notes that production systems must use the enclave binary from the latest release, as builds are not reproducible since they are signed with an undisclosed key.
We built version \texttt{v1.14.0} in a Docker container and, as expected, failed to reproduce the measurement.
The developers confirmed that builds are non-reproducible and that the production enclave is indeed signed.
As the inclusion of the signature is a design feature, we conclude that the Secret Network is non-reproducible by design. In other words, the lack of reproducibility prevents users from independently verifying that the production enclave runs the expected code and is free of vulnerabilities or backdoors.

This limitation has important implications: although the source code is publicly available, users cannot verify that the deployed enclave corresponds to the audited code. As a result, trust in the system ultimately depends on the developers, weakening the intended transparency guarantees.

Despite this limitation, the platform is actively used in practice. 
While the SGX-based Secret Network node (also available on the Azure marketplace) lacks reproducibility, a newer version leveraging AMD SEV-SNP and Intel TDX now offers reproducibility support, including build and verification instructions~\cite{secret_network_full_verification}, indicating a stronger emphasis on reproducibility in recent TEE deployments.

Following these instructions, we successfully reproduced the SEV-SNP measurement for Secret VM \texttt{v0.0.25}. %
For TDX, reproducibility was only partially achieved. Recall that TDX exposes four additional measurement registers alongside the initial VM measurement at boot (cf. \Cref{sec:bg_tees}). While the initial VM measurement and one of the additional registers matched the values retrieved from the release, the remaining three did not.  
This discrepancy may be due to incomplete instructions for retrieving the measurements, which required modifications to the command invoking the suggested measurement tool. However, since the same command was used for both the released and locally built artifacts, this explanation is unlikely. At the time of writing, the issue remains unresolved, as we did not receive a response from the developers.

Similar patterns appear in other SGX-based systems. 
For instance, based on communication with their developers, neither Nginx-SGX~\cite{nginxsgx_repo} nor MongoDB-SGX~\cite{mongodbsgx_repo} is reproducible. Both are enclavized versions of NGINX and MongoDB built with Gramine~\cite{gramine_repo}, and both take deliberate shortcuts that favor ease of deployment over reproducibility. Rather than injecting certificates after successful remote attestation, they generate self-signed certificates dynamically at build time. Because these certificates differ from build to build yet are covered by the enclave measurement, the measurement itself cannot be reproduced.

\paragraph{CCF}
\label{sec:ccf}

The Confidential Consortium Framework (CCF)~\cite{russinovich2019ccf} is an open-source framework for high-performance decentralized applications, providing confidentiality, integrity, and availability via a TEE-based consensus protocol. Starting with version \texttt{6.0.0}, CCF discontinued support for Intel SGX and now exclusively targets AMD SEV-SNP. However, the Azure service ``Confidential ledger''~\cite{ccf_azure} still utilizes the SGX-based CCF in the backend~\cite{ccf_repo_sgx}. 
We successfully compiled CCF \texttt{v5.0.6}, but the enclave measurement did not match the release. Achieving reproducibility required the CI/CD build path (\texttt{/\_\_w/CCF}), which was kindly provided by the developers.
With the transition from Intel SGX to AMD SEV-SNP, Microsoft appears to have prioritized reproducibility, similar to the Secret Network. The updated documentation includes a dedicated section on reproducible builds~\cite{ccf_reproducibility_doc}, providing sufficient information to achieve reproducibility without developer assistance. 

\paragraph{Gramine and Ternoa}
\label{sec:gramine_ternoa}

Similar to CCF, reproducing Gramine~\cite{gramine_repo} and Ternoa~\cite{ternoa_website} was hindered by incomplete build environments.

For Gramine \texttt{v1.8}, building on the documented platform (Ubuntu 22.04) produced enclave measurements that did not match the official release. Reproducibility was only achieved by replicating the full release environment, including Debian 11 (with backports), exact dependencies, compiler flags, and environment variables provided by the developers.

For Ternoa, building enclave \texttt{v0.4.5-mainnet} failed due to unresolved transitive dependencies. Although we obtained a working build after manually fixing multiple issues, the resulting binary did not match the release. According to the developers, some dependencies required for full reproducibility remain unknown, making reproduction infeasible. As with CCF and Gramine, the issue was insufficient build documentation; however, the missing information could not be recovered in this case.

\paragraph{MobileCoin and Mystikos}

For eleven of the 17 unreproducible projects, the cause remains unknown: unlike the Secret Network and Gramine cases, we received no response from the developers. Because enclave measurements are one-way hashes over every bit of the binary, we can only \emph{speculate} about the source of non-determinism. Some causes—timestamps, build paths, file systems—are easy to spot; others, such as a mismatched dependency version, are far harder to pin down.

Given how individualized the remaining build processes are, we selected two projects for manual analysis in order to rule out candidate causes: MobileCoin~\cite{mobilecoin_repo}, a privacy-preserving payment network, and Mystikos~\cite{mystikos_repo}, an SGX runtime for Linux applications. Both builds turned out to be deterministic and independent of build time. The Mystikos build, however, depended on the build path. We recovered the original path from the published binary and rebuilt with it, yet the result still did not reproduce, so at least one further factor must influence the output. Varying the file system across ext4, xfs, and btrfs left MobileCoin's measurement unchanged, but altered the Mystikos binary hash between ext4 and xfs/btrfs. Recompiling Mystikos with the corrected build path produced yet another hash. Together, these results illustrate how hard it is to identify the root cause of non-reproducibility without access to the original build environment.

\begin{solution}
As shown earlier, our analysis of the case studies did not reveal any technical challenges that were inherently difficult to overcome. Most issues could be addressed by adding more detailed information to the documentation, suggesting that fundamental technical limitations are not the main obstacle. In the following section, we explore the underlying reasons for the lack of reproducible TEE builds in practice through an interview study.
\end{solution}

\section{Interview Study}
\label{sec:interview_study}

Having established that reproducibility challenges extend across SGX, TDX, and SEV-SNP, we turn to understanding the deeper factors at play behind these challenges. To this end, we conducted an exploratory qualitative interview study with 12 maintainers of open-source SGX projects from our dataset. All participants had several years of experience developing TEE software, primarily for SGX, with some also having experience with other TEE platforms. We focused on Intel SGX given its dominant ecosystem and comparatively mature developer base, while also inviting participants to reflect on their experiences with other TEEs where relevant. Our aim was to investigate developer practices, surface challenges, and identify opportunities for improvement in the reproducibility of TEE-based build pipelines.
In what follows, we present our methodology and the results of the interviews. 

\subsection{Methodology}
\label{sec:study_methodology}

\begin{figure}[t]
    \centering
    \includegraphics[width=0.98\linewidth]{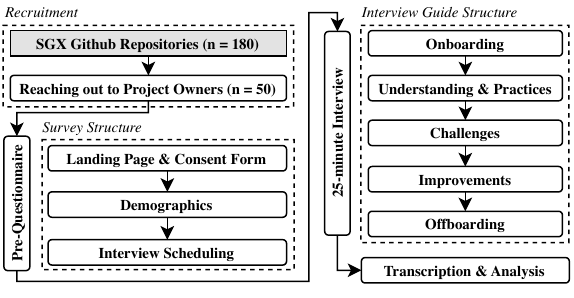}
    \caption{Overview of our methodology.}
    \label{fig:methodology}
\end{figure}

To understand how our target group approaches reproducibility in the context of SGX, we adopt a qualitative, exploratory research design based on semi-structured interviews. This approach is appropriate given the relatively small size of the population and the suitability of interviews for gaining in-depth insights into individual practices, processes, structures, and perceptions. Based on the repositories identified in~\cref{sec:reproducibility_in_the_wild}, we reached out to potential participants to discuss their development practices. To collect demographic information in a structured way, save time during the interviews, and simplify scheduling, we distributed a short pre-questionnaire prior to the 25-minute interviews. \Cref{fig:methodology} provides an overview of our methodology and its main components.

\subsubsection{Recruiting.}
We reached out by email to the contributors of all SGX GitHub projects identified in~\cref{sec:reproducibility_in_the_wild}, where contact information was available. In our recruitment email (cf. 
\Cref{fig:interview_email} in Appendix~\ref{apx:codebook}), 
we transparently explained that we are researchers seeking to advance TEE security in the context of reproducibility. We explicitly noted that no prior experience with reproducibility was required to participate. Participants were offered a €50 gift card (5000+ digital reward options in 170+ countries) as compensation and could schedule an interview directly at the end of the pre-questionnaire. If no response was received after the initial contact, we sent one polite reminder and then refrained from further follow-ups. Three participants chose to waive their compensation in the pre-questionnaire.

\subsubsection{Piloting.}
The interview guide (cf. \Cref{apx:interviewguide}) 
was developed deductive-iteratively. Four researchers---two with technical expertise and two with a background in human-centered security---inde-pendently proposed questions based on the earlier identified reproducibility issues (cf.~\cref{sec:reproducibility_in_the_wild}). The questions were then collaboratively selected, grouped, and refined. The guide was piloted twice internally and twice externally with participants (P1 and P2) from our professional network, who worked with SGX in the past. The most substantial change resulting from the pilot concerned the definition of reproducibility in the context of SGX, which we refined to ensure a consistent understanding among participants for the subsequent interview questions. The pilot participants are therefore not included in our analysis or results.

\begin{table}[t]
	\caption{Summary of identified reproducibility blockers, including excluded sources of non-determinism for two projects without a developer response.}
	\label{tab:root_causes}
	\centering

	\scalebox{0.86}{
    \footnotesize
	\begin{tabular}{ |l|l|p{0.4\linewidth}| }
		\hline
		  \textbf{Project} & \textbf{Reproducibility blockers} & \textbf{Excluded root causes} \\
		\hline
        
        \rowcolor{gray!10}
        \multicolumn{3}{|l|}{\bf Reproducible in guided scenario} \\
        \hline 
		CCF \cite{ccf_repo_sgx} & Build path & -- \\
		Gramine \cite{gramine_repo} & Undisclosed dependencies & -- \\
		Marblerun \cite{marblerun_repo} & Undisclosed build configuration & -- \\
        
        \hline 
        \rowcolor{gray!10}
        \multicolumn{3}{|l|}{\bf Unreproducible in guided scenario} \\
        \hline
		MongoDB-SGX \cite{mongodbsgx_repo} & Self-signed certificate & --\\
		Nginx-SGX \cite{nginxsgx_repo} & Self-signed certificate & --\\
		Safeheron Arweave \cite{safeheron_repo} & Unavailable dependency & -- \\
		Secret Network \cite{secret_network_repo_sgx} & Embedded secret & -- \\
		SGXSSE \cite{sgxsse_repo} & Unavailable dependency & -- \\
		Ternoa enclaves \cite{ternoa_enclaves_repo} & Unknown dependencies & -- \\
        \hline 
        
        \hline 
        \rowcolor{gray!10}
        \multicolumn{3}{|l|}{\bf Unreproducible and no guided scenario (no developer response)} \\
        \hline 
		MobileCoin \cite{mobilecoin_repo} & -- & Compiler randomness, build path, file system, timestamp \\
		Mystikos \cite{mystikos_repo} & Build path, file system & Compiler randomness, timestamp  \\

		\hline

	\end{tabular}}
\end{table}

\subsubsection{Participants.}
From 180 SGX GitHub projects, we identified contact information for 67 contributors across 55 projects. Among those, the contact information of five projects was outdated. Of the 62 email invitations sent, we received 13 responses, and eleven participants agreed to schedule an interview. One participant, due to language difficulties, asked to communicate in writing rather than verbally. We accepted this mode of participation, meaning 12 interviews form our final dataset. Our participants came from eleven different countries, with an average age of 33.08 years and almost 16 years of programming experience. In terms of their roles, half of the participants hold academic positions, and the others work in industry. In line with common practice,  we only report participants' aggregated demographic information to prevent re-identification. \Cref{tab:demographics} summarizes these aggregated demographics, including an overview of the participants' positions and roles.

\subsubsection{Interview Structure \& Procedure.}
The semi-structured interviews were conducted with two researchers present. One served as the interviewer, while the other ensured adherence to the interview guide and handled technical aspects, with roles alternating across sessions. The researcher in the assistant role introduced themselves at the beginning, but then turned off their video and remained silent to avoid distracting the participant.

The interview guide is structured into the five main areas:
\textcircled{1} \emph{Onboarding}: During onboarding, the researchers briefly introduced themselves, the institution, and the purpose of the study to build a rapport with the participants. We asked if they had any questions about the consent form, and asked them to agree again to the interview being recorded.
\textcircled{2} \emph{Understanding \& Practices}: We first asked what reproducibility means for the participants, capturing their individual perception of the concept and the perceived importance. If a participant's interpretation diverged from the study definition, the interviewer reiterated our definition of reproducibility developed following the pilot interviews to ensure a consistent baseline. The participants then explained the extent to which reproducibility was relevant in their current project.
\textcircled{3} \emph{Challenges}: We followed up on previously mentioned issues and explicitly invited participants to discuss any additional challenges or problems they had encountered in their own projects, as well as those related to other TEEs.
\textcircled{4} \emph{Improvements}: We asked participants how the reproducibility challenges in SGX and other TEEs could be addressed, and discussed appropriate channels for disseminating reproducibility-related information and ways the community could strengthen reproducibility practices.
\textcircled{5} \emph{Offboarding}: We thanked the participants for taking part, gave them the opportunity to ask final questions, and outlined next steps, including issuance of the voucher and the interview evaluation process.

\paragraph{Analysis}
All interviews were recorded and transcribed locally using Whisper~\cite{whisper}, then manually checked for accuracy, and potentially identifying information was redacted where necessary. We conducted the analysis following the principles of thematic analysis~\cite{clarke2014thematic}. After each interview, the two researchers who conducted the interview sessions exchanged first impressions and took initial notes. Subsequently, the research team held joint analysis sessions to deepen their understanding of the data, clarify open questions, and establish an initial version of the codebook. %
Following the joint session, one researcher coded the data and iteratively refined the codebook as new aspects emerged. 
After finalizing the codebook and the initial coding, a second researcher reviewed all codes and discrepancies were resolved through discussion.
The resulting codebook was used as a shared analytical framework for the subsequent steps and is included in the appendix (cf. \cref{apx:codebook}).
In several iterative discussions within the research team, we used the refined codes to synthesize higher-level themes and structure the presentation of results. These sessions also informed the development of overarching themes and insights discussed in~\cref{sec:discussion}. 

\subsection{Results}
\label{sec:interview_results}

We now present the interview results, following the structure of our codebook (cf. \Cref{tab:codebook-tee-reprod} in Appendix \ref{apx:codebook}). We begin with participants' perceptions of reproducibility, followed by the associated challenges and how reproducibility is prioritized within their projects. We then discuss use cases and enablers of reproducibility, and conclude with current practices and suggested improvements.

\subsubsection{Perception of Reproducibility.}
Participants discussed two areas: the importance of reproducibility for TEEs and what reproducibility meant to them. Afterward, the researchers provided their definition of reproducibility to ensure a shared understanding of the concept.

\paragraph{Relevance of reproducibility} 
Eight participants commented on the relevance of reproducibility. Five participants stated that it is essential for ensuring security, and three participants stated that a lack of reproducibility directly leads to a lack of security: \pquote{It's not a feature. It's an assumption. So if this assumption breaks, then our system breaks.}{P7}

However, two participants also considered reproducibility irrelevant for end-users of products, as it might be challenging for them to test reproducibility: \pquote{You want to enable people to do it themselves, but only a few people will do it themselves, right? Most people will want to trust some service or some statement or some public value. And doing this transparently is also difficult, right.}{P9}

\paragraph{Definition of reproducibility} 
We identified three different patterns about reproducibility in the context of SGX, highlighting the lack of a consistent understanding of what reproducibility actually entails. Participants also gave us multiple ideas and definitions themselves. For seven participants, ``reproducible'' meant publishing results or application output that could be replicated: \pquote{Once a research or a group of researchers publish their work, other researchers around the world are able to get similar results by running, by executing, by reproducing exactly what these researchers made available.}{P8}. Three participants described reproducibility as creating the same binary from the source code, which was very close to our definition of reproducibility. Five associated the notion of reproducibility with applications \simplequote{working as expected}. We further asked participants in what context they had learned about reproducibility. Although not all participants responded (9/12), four said they had gained their knowledge in an academic setting and three described it as a general understanding. Only one participant said they had learned about it in an educational setting or in the workplace. %

\begin{table}[t]
    \caption{Demographics of the participants (N=12).}
    \label{tab:demographics}
    \centering
    \scalebox{0.95}{
    \begin{threeparttable}
        \footnotesize
        \renewcommand{\arraystretch}{1.05}
        \setlength{\tabcolsep}{0.3\tabcolsep}
        \setlength{\defaultaddspace}{0.08\defaultaddspace} %

        \begin{tabular}{|@{\hspace{0.4em}}lrr@{\hspace{0.4em}}|@{\hspace{0.4em}}lrr@{\hspace{0.4em}}|}
            \hline

            \rowcolor{gray!10}
            \multicolumn{6}{|@{\hspace{0.4em}}l|}{\textbf{Gender}}  \\ \hline
            Male & 12 & 100\% &  &  &  \\ \hline

            \rowcolor{gray!10}
            \multicolumn{6}{|@{\hspace{0.4em}}l|}{\textbf{Age [years]}} \\ \hline
            Min. & 26.00 &  & Max. & 40.00 & \\
            Mean, Std. & 33.08 & $\pm4.44$ & Median & 33.00 & \\ \hline

            \rowcolor{gray!10}
            \multicolumn{6}{|@{\hspace{0.4em}}l|}{\textbf{Highest Education Level}}  \\ \hline
            Bachelor's degree\hspace{-3em} & 1 & 8\% & Master's degree & 5 & 38\% \\
            Doctoral degree (PhD) & 6 & 46\% &  &  &  \\ \hline

            \rowcolor{gray!10}
            \multicolumn{6}{|@{\hspace{0.4em}}l|}{\textbf{Country of Residency}}  \\ \hline
            Germany & 2 & 15\% & Brazil & 1 & 8\% \\
            China & 1 & 8\% & India & 1 & 8\% \\
            Japan & 1 & 8\% & Singapore & 1 & 8\% \\
            Slovenia & 1 & 8\% & Sweden & 1 & 8\% \\
            Switzerland & 1 & 8\% & Thailand & 1 & 8\% \\
            UK & 1 & 8\% &  &  &  \\ \hline

            \rowcolor{gray!10}
            \multicolumn{6}{|@{\hspace{0.4em}}l|}{\textbf{Coding Experience [years]}}\\  \hline
            Total Median & 15.00 &  & Prof. Median & 6.00 & \\
            Total Mean, Std. & 15.92 & $\pm5.32$ & Prof. Mean, Std. & 7.08 & $\pm4.25$ \\
            Total Min. & 9.00 &  & Prof. Min. & 2.00 & \\
            Total Max. & 25.00 &  & Prof. Max. & 18.00 & \\ \hline

            \rowcolor{gray!10}
            \multicolumn{6}{|@{\hspace{0.4em}}l|}{\textbf{Weekly Dev. Time [hours]}}\\  \hline
            less than 5 hours & 4 & 31\% & 10--20 hours & 3 & 23\% \\
            20--30 hours & 4 & 31\% & 30--40 hours & 1 & 8\% \\ \hline

            \rowcolor{gray!10}
            \multicolumn{6}{|@{\hspace{0.4em}}l|}{\textbf{Employment Status}}\\  \hline
             Employed full-time & 10 & 77\% & Self-employed & 1 & 8\% \\
             I am not gainfully employed. & 1 & 8\% &  &  &  \\

            \hline

            \rowcolor{gray!10}
            \multicolumn{6}{|@{\hspace{0.4em}}l|}{\textbf{Position/role}}\\  \hline
            Academic researcher/scientist & & & P3, P5, P6, P7, P8, P11, P12 & & \\
            Developer & & & P4, P6, P12 & & \\
            Engineer & & & P4, P14 & & \\
            CEO/Manager/Team lead & & & P4 & & \\
            Educator & & & P3, P6, P11 & & \\
            Security specialist & & & P4, P9, P10, P13, P14 & & \\

            \hline
        \end{tabular}
    \end{threeparttable}}%
\end{table}

\subsubsection{Challenges.}
\label{sec:interview_challenges}
Participants reported various challenges they encountered within their project, or which they perceived as challenging. While most participants referred to SGX, four of the 12 noted that the challenges are often not specific to a given TEE, but apply to confidential computing in general. An overview of the participants' mentioned challenges is presented in~\cref{tab:reproducibility_challenges} (here we also highlight which of these challenges we encountered---either fully or partially---during our own analysis in~\Cref{subsec:pitfalls}).

\paragraph{Limited control over the environment}
The challenge that was most frequently cited by participants (6/12) was limited control over the environment. This included the impact of dependencies on reproducibility, the complexity and dynamics of the development environment, and the fact that the build environment often does not support reproducibility. P12 explained the complexity of considering all parts that might impact the hash of a binary: \simplequote{In my experience a lot of moving parts, system or program which is complex enough has so many moving parts, it's pretty hard to count them all in with at least regular tools.}. P4 also elaborated on the challenges of distributed build environments: \pquote{We're aware that for bigger build systems, you might have a situation where you're distributing the build across machines and then maybe the optimizations or the, the final binary that's produced depend on how the distribution has been done.}{P4} P6 further added, that in some project one does not have control over the build environment, as they might not be maintained by the development team. Dependencies might be integrated in the environment (e.g., in Ubuntu itself), they further elaborated on their concern, that \simplequote{the developers have to be very precise about the dependency they're using, make sure that they're coming from the same source, compiling using the same tools, etc.} 

Five other participants also mentioned challenges related to software dependencies. P3, P4, and P6, for example, explained that major changes to packages can also affect the compiler tool set, and that everything must be managed accordingly. In the same context, P3 mentioned that they used multiple build environments: \simplequote{Actually, for that, I think in one of my earlier projects, I just like have multiple gcc versions. So for different enclaves, I need to, like, switch to that version, and then switch back.}. It was also noted that dependencies can change rapidly over time and must be managed accordingly. Legacy software and packages that are no longer maintained also pose a challenge. Software packages may no longer be available, or patches may need to be applied manually, which can consequently affect the build process or the build environment(s). 
One participant also criticized the compilers' lack of support for reproducible builds: \pquote{But it's more like the tooling. The tooling is not aware of this. And I don't know how aware the compiler writers are of the importance of maintaining reproducibility. They could make optimization decisions that would make the builds faster or more efficient in some way, but at the expense of reproducibility.}{P4}

\paragraph{Build-induced irregularities}
Irregularities introduced during the build process were mentioned by five participants. Four participants mentioned that timestamps left in the build binary \simplequote{are a really common issue}. Paths embedded during the compilation process also posed a challenge for two participants. Related to paths introducing irregularities, P4 stated: \simplequote{Another thing is that anything that sets symbols in the binaries tends to capture the absolute path where you run the builds. For SGX, we tended to kind of sidestep that.}

\paragraph{Lack of awareness}

Four participants either directly stated that they or others are not aware of the importance of reproducibility in the context of the security of SGX applications, or they had misconceptions about it. For example, one participant assumed that reproducibility depends primarily on their own code rather than on the \simplequote{underlying dependencies}, or it was pointed out that researchers may think of artifact evaluation instead when it comes to reproducibility.

\paragraph{High complexity}
Four participants stated that the complexity of the concepts behind secure enclaves makes reproducibility difficult. P10, for example, pointed out that SGX development is \simplequote{notoriously complex} and that this \simplequote{discourages developers from focusing on reproducibility in already highly pressured environments}. P14 also emphasized the complexity of achieving reproducibility by stating that they knew people at a large tech enterprise who believed reproducibility could not be achieved within the available budget. P10 specifically explained that reproducibility becomes even more challenging in the future: \simplequote{With newer CVM technologies like TDX or SEV-SNP, the enlarged TCB makes reproducibility virtually impossible to guarantee. Attempting to ensure reproducibility across such massive stacks is clearly infeasible, further shifting attention away from reproducibility.}. 

\paragraph{Debugging non-reproducibility is hard}
Three participants also reported that compilation alone, without accounting for reproducibility, posed a major challenge in general. Two participants reported that debugging was a significant challenge when reproducibility failed. Factors contributing to this included insufficient documentation of the components that influence the hash (P10) and difficulty in identifying sources of non-determinism (P14).

\paragraph{Non-deterministic signatures \& hotfix release before fix}
One participant (P4) mentioned that generating and embedding the signature early in the build process makes it non-deterministic and that the rapid release of the source code after a hotfix---while necessary for reproducibility---also allows attackers to exploit vulnerabilities before end users have applied the patch.

\begin{table}[tbp]
\centering
\footnotesize
\caption{Overview of challenges reported by the participants. The dots indicate whether we encountered the challenges in our technical analysis completely, partially, or not at all.}
\label{tab:reproducibility_challenges}
\begin{tabular}{|l |c| c|}
\hline
\textbf{Challenge} & \textbf{\# P.} & \textbf{Encountered} \\ 
\hline
Limited control of the environment & 6 & \yes \\ 
Lack of Awareness & 4 & \no \\
High complexity & 4 & \no \\
Debugging non-reproducibility is hard & 2 & \may \\
Challenges are often not TEE specific & 2 & \yes \\
Non-deterministic signatures & 1 & \yes \\ 
Timestamps in Binaries & 3 & \no \\
Path resolution & 2 & \yes \\
Hotfix release before fix & 1 & \no \\
\hline
\end{tabular}
\end{table}

\subsubsection{Prioritization of Reproducibility.}
\label{sec:priority}
We found that only one participant (P14) perceived reproducibility as a priority in their project. The participant 
described it as \simplequote{one of our major goals} with non-reproducibility being a \simplequote{release blocker}. On the other hand, eight participants reported that reproducibility had little or no priority in their projects. One participant explicitly elaborated on the situation of projects from an academic context: \pquote{This intense academic competition meant that issues like reproducibility---although important---were often de-prioritized if they did not immediately lead to security breakage.}{P10}. They continued elaborating on commercial products: \pquote{On the market side, products such as PowerDVD or blockchain-related solutions did appear. However, here too, due to commercial competition or the involvement of maintainers who may not be deeply familiar with TEEs, specialized issues like reproducibility were again often neglected.}{P10}. P6 reflected on their own experience, where they prioritized other tasks over achieving reproducibility: \pquote{I think that there are things that are a bigger priority right now. For example, when I was trying to, when I was developing my own SGX software, there was a big issue with attestation because it's really difficult to use [...] I think that there are a list of priorities in SGX deployments and we are kind of working down the list. We still haven't gone past some really important issues, like even to get things to compile, even to get the framework to stabilize and usable and user-friendly.}{P6}. They also explained that companies \simplequote{know this is a problem, but they don't care about this} and that they \simplequote{don't want to invest manpower}. In contrast to the other participants, reproducibility was a priority for P14, but they also stated that customers might not value it: \pquote{For us, it's a selling feature that our stuff is reproducible, and you can validate it. But yeah, the usually contact to a customer is like, okay, they think it is really important and then they learn about what it means to verify it. And then they are, maybe it's not that important. It's enough if you can like have a checkbox and yeah.}{P14}

\subsubsection{Practices to achieve reproducibility.}
In the interviews, we also discussed our participants' experiences who had addressed, or attempted to address, reproducibility. As outlined in~\cref{sec:priority}, reproducibility was not a priority during development for most participants. However, two participants elaborated in more detail about the time they invested in achieving reproducibility. P4, for example, explained that reproducibility had not been considered from the beginning of the project and that they later assigned one full-time employee to address it: \pquote{I think [the employee] spent about five to six weeks, I would say, improving the reproducibility of the process, making it path independent, having a CI job that was doing that, producing a script so that end users could just run the script with a metadata file and reproduce things.}{P4}. They continued explaining that \simplequote{reproducibility is not necessarily something you have straight away}. But after they achieved reproducibility, they also made clear that they do expect to invest much effort in the future. 
P14 also stated that, depending on their situation, they spent a considerable amount of time on reproducibility, but in general it did not take much from the development budget: \pquote{Maybe 5\% to 10\% over the years. But that that greatly differs what we are working on, right? So we had times where we spent like full months on reproducibility.}{P14}. 
P7 guessed, that for commercial projects the effort should remain below 30\%. 

Eight participants further elaborated on tools they use or consider to use for achieving reproducibility. All of them mentioned Docker. While four were already using Docker, and four reported that Docker is a valuable approach for achieving reproducibility, it was also criticized. P14 noted that \simplequote{The inputs are not pinned. So if you do an apt install or an update in the Docker container, you will just get anything}. Other tools named as valuable were Nix (2), Bazel (1), and the Yocto Project (1). Specifically, Nix and Bazel are described as useful for achieving reproducibility, because they require developers to be more strict about their input specifications: \pquote{Nix has quite a strong sandboxing. So we know that all the inputs that we require are pinned by hash because that's enforced by Nix, actually. [...] We also use Bazel in some other projects, but the concept is the same, right? You need to pin all your inputs by hash.}{P14}. 

Four participants also commented on the practices they used during the build process to achieve reproducibility. The focus was particularly on the need to pre-define sources for non-determinism. This included that participants used a \pquote{single machine}{P4} or the same \pquote{environment for a couple of years, so it's kind of stable}{P3}. The use of static build paths, fixed timestamps, or fixed entropies for signatures was also mentioned by individual participants. 

Among the participants, two elaborated on the sources from which they learned how to make applications reproducible. While both participants stated that trial and error was part of the learning process, P4 also pointed out that \simplequote{a lot of the techniques and the tooling come from Linux distro people} and that they are unaware of a single source of reference for reproducibility: \pquote{It was very much a sort of look, look for things, look for techniques. I don't think we have a single, single source of reference.}{P4}.

\subsubsection{Suggested improvements.}
All but one participant provided suggestions on how to improve reproducibility. Their ideas included both technical improvements and general recommendations, as well as specific actions researchers could take and channels through which information about reproducibility should be disseminated. 

\paragraph{General desirable improvements} Four participants demanded that \pquote{we should increase awareness.}{P7}. P7 further elaborated stakeholders are unlikely to prioritize reproducibility without a triggering security incident: \pquote{So yeah, it's not easy. So unless, so if they see some problem from reproducibility, so they will invest efforts on this. So yeah, we need to maybe try to provide some kind of like attack. It's not real attack, but we try to present the problem to the industry to let them know that this is a big problem. [...] So the very big zero-day vulnerability, so the company will pay attention very quickly.}{P7}. P10 also highlighted the importance of raising awareness: \pquote{Perhaps most importantly, raising awareness of reproducibility as a critical issue will itself have a major impact.}{P10}

Other general improvements included, for example, \pquote{extending version-managed builds like Yocto Project}{P10}, building reproducibility tools in GitHub, for example, \pquote{some kind of script that checks if it's following all the reproducible guidelines}{P8}, or developing tools that simplify debugging (P14).

\paragraph{Technical improvements} One technical improvement mentioned was that the entire compilation environment could be built and offered in Docker (P6, P7). Other suggestions were the use of SBOMs to track the tools used in the environment (P3, P9), introducing reproducibility features in package managers (P7), or a trusted authority issuing code-signing certificates for enclave code (P10).

\paragraph{Support through research}
Six participants expressed wishes for how research could support the dissemination of reproducible builds through appropriate projects. P5 and P8 asked for specific best practices for achieving reproducibility that could be defined by research. P4 suggested creating sample projects in which tools could be evaluated and reproducibility practices could be traced. P10 hoped that \simplequote{having researchers directly join projects as maintainers could improve reproducibility quality}. P14 also suggested research teams could evaluate the reproducibility of different ecosystems.

\paragraph{Information channels} Six participants elaborated on the channels they preferred for information on SGX reproducibility. Overall, however, there seemed to be no consensus on this point. While three participants mentioned the Intel website, other channels, such as scientific conferences, attestation services, and GitHub repositories, were discussed only by individual participants.

\section{Discussion}
\label{sec:discussion}

We now discuss and interpret the results from our analysis of TEE projects presented in~\cref{sec:reproducibility_in_the_wild} and the interview study described in~\cref{sec:interview_study}. Table~\ref{tab:codebook-tee-reprod} in the appendix maps the themes discussed in this section to the corresponding codes of our codebook.

Our analysis revealed a general lack of reproducibility among open-source TEE projects. Specifically, only ten of the 115 analyzed projects, i.e., 8.7\%, were reproducible (cf.~\cref{sec:technical_results}). Among the three projects that required additional developer input to achieve reproducibility, the lack of relevant documentation was the primary reason (cf.~\cref{subsec:pitfalls}); once the missing information was obtained, achieving reproducibility was straightforward. Nonetheless, this raises the question of why reproducibility support remains so scarce, given that the ability to reproduce enclave binaries is an important component for remote attestation---and therefore essential for the secure deployment of TEEs.

\subsection{Trends Across TEE Technologies}
\label{sec:discussion_other_tees}

Our paper examined three TEE technologies: Intel SGX, Intel TDX, and AMD SEV-SNP. Among these, SGX is the oldest and most extensively studied, TDX is the most recent, and SEV-SNP was released in between. All three ground their security guarantees in remote attestation, for which reproducible builds are a key prerequisite to enable independent verification of enclave measurements. Arm TrustZone, by contrast, does not natively support remote attestation, making build reproducibility less critical in that context---though it remains beneficial for mitigating supply-chain attacks.

Our technical analysis of 115 TEE deployments finds no statistically significant association between the TEE technology and reproducibility (cf. \Cref{sec:technical_results}). This finding is reinforced by our interview study, in which participants reported similar challenges across platforms, including non-deterministic build inputs, insufficient documentation, and complex dependency management (cf.~\Cref{sec:interview_challenges}).
Taken together, these results suggest that reproducibility challenges are TEE-agnostic and that our findings (along with the broader implications discussed below) generalize across TEE architectures and likely extend to future TEE designs that similarly rely on remote attestation and enclave measurement verification.

\subsection{Reproducibility Challenges}
\label{dis:Challenges}

\paragraph{Developers' Challenges: Known Issues}
Our interview study revealed several technical issues that complicate realizing reproducibility in practice (cf.~\Cref{sec:interview_challenges}), many of which are known from prior work on traditional reproducible builds~\cite{DBLP:journals/software/LambZ22/rbTechnicalChallenges}. Five participants, for instance, mentioned timestamps included in the binary---introduced via macros (e.g., C's \texttt{\_\_DATE\_\_}) or build tools---as a common source of non-determinism. 
Similarly, divergent file system ordering across hosts can alter processing order, and absolute file paths embedded in binaries (e.g., via the \texttt{\_\_FILE\_\_} pre-processor macro) vary with the build directory, eventually leading to non-identical artifacts. We also observed the latter in our analysis of an SGX project (cf.~\Cref{sec:ccf}).
Additionally, randomness from undisclosed secrets and non-deterministic signatures was identified as a major obstacle, consistent with our empirical findings (cf.~\Cref{sec:secret_network}).
Finally, dependency management emerged as a particularly challenging aspect in the interviews. This is reflected in our empirical results, which show that tools such as Nix~\cite{nix}, Bazel~\cite{bazel}, and Yocto~\cite{yocto}---all three enforcing externally defined and versioned build inputs (e.g., dependencies)---substantially improve reproducibility (cf.~\Cref{sec:technical_results}).

\paragraph{Organizational Barriers: Problems Exceed the Developers' Responsibility}
A key reason identified in our interviews for the lack of practical reproducibility support---beyond known technical challenges---is developers' limited control over the build environment (cf.~\cref{sec:interview_challenges}). In many projects, builds are executed in CI/CD\footnote{Continuous Integration and Continuous Deployment} pipelines that are %
centrally managed by DevOps teams or shared across multiple teams, rather than controlled by individual developers. As a result, even straightforward technical changes can become significant organizational hurdles, hindering the adoption of deterministic build practices.
Moreover, software development is typically collaborative, often spanning multiple teams or organizations. If any component in this chain fails to ensure reproducibility, the entire build process may become non-deterministic. One participant noted that a single non-reproducible dependency can compromise the entire application's reproducibility (cf.~\cref{sec:interview_challenges}). Addressing such issues typically requires coordination with external parties or manual tracing of non-determinism---both time-consuming efforts: in our empirical analysis, only 35\% of contacted development teams responded, and in four cases, this information was still insufficient to reproduce the build (cf. \Cref{sec:technical_results}). Similarly, participants emphasized that debugging non-determinism in large, dependency-rich code bases is complex and resource-intensive.

These findings suggest that practically improving reproducibility in TEE-based applications requires more than technical knowledge among developers.

\subsection{Security Is A Badge}
\label{sec:discussion_badge}

Recall that security is a central feature of TEE projects, with remote attestation forming a core component of this model.
Eight interview participants consider reproducibility highly important---or even as a prerequisite---for ensuring security in SGX-based systems, confirming this perspective.
Despite this, our empirical findings show that reproducibility is rarely achieved in practice. %
This aligns with feedback from all but two participants who noted that reproducibility is generally not treated as a development priority (cf.~\cref{sec:priority}). 
\emph{These results indicate a disconnect between the perceived importance of reproducibility and its prioritization in practice, driven by decision-making rather than developer awareness.}

Over a project's lifecycle, evolving organizational structures, increasing dependencies, and personnel turnover make retrofitting reproducibility increasingly difficult. One participant reported that achieving reproducibility required five to six weeks of dedicated developer effort once it was prioritized---an investment many organizations avoid without clear incentives. Prior work similarly shows the difficulty of dependency discovery, even with specialized tools~\cite{Cofano.2024, Stalnaker.2024}.
\emph{This highlights the need for stronger organizational and economic incentives to address reproducibility early. }

Given that remote attestation has limited value without reproducible builds, a key question arises: \emph{how do customers and stakeholders perceive the value of attestation, and is the mere label ``uses TEE'' considered sufficient?} Without a technical understanding of attestation, customers may not demand reproducibility, thereby weakening incentives for companies to invest in it. Similar patterns appear in other security domains, where the cost of implementing robust protections often outweighs the perceived financial risk of incidents~\cite{10.1145/581271.581274}. Consistent with this, one interview participant noted that reproducibility would only become a priority if customers explicitly demanded end-to-end attestation support (cf.~\cref{sec:priority}).

\subsection{Study Limitations}
\label{sec:limitation}

Although we contacted developers for all 50 out of 180 SGX projects for which we could locate contact information, only 12 agreed to be interviewed, and we were unable to reach additional developers. 
Reasons for this limited response can range from a lack of time and legal constraints to a lack of prioritization of reproducibility. Nonetheless, we note that 71\% of the development teams contacted for our technical study were unresponsive (cf. \Cref{tab:study}), suggesting that support for aiding reproducibility could be improved. While the response rate limited the sample size of our interview study, it covers diverse participants' backgrounds and TEE expertise, capturing a range of perspectives, and thus offers exploratory evidence about the perception of reproducibility among TEE developers and the barriers they face.
Note that nearly half of the participants were from academia, where reproducibility is often not a primary objective and enclave implementations are typically prototypes that may not support remote attestation. Nonetheless, our technical analysis---including production-grade industry projects---shows that poor reproducibility is equally prevalent in industrial-scale TEE deployments. 

Finally, our interview-based findings reflect the usual constraints of qualitative self-reported data, including recall bias and social desirability bias, particularly among maintainers who may be reluctant to portray their projects or organizational practices negatively.

\section{Recommendations}
~\label{sec:recommendations}

Based on our technical analysis and expert interviews, we recommend that TEE stakeholders raise awareness of reproducibility, provide practical guidance, and establish it as a foundational concern in TEE development. 

\subsection{Application Developers}
To improve reproducibility without disrupting established development workflows, we propose the following recommendations for application developers:
\begin{enumerate}[leftmargin=*]
    \item \textbf{Increase awareness and understanding of reproducibility.} Our interviews reveal limited consensus on reproducibility concepts, with 33\% of developers unaware of its importance for TEEs and many reporting diverging interpretations shaped by academic or general software perspectives. We therefore recommend standardized educational materials that explain the importance of reproducibility for trusted execution and account for developers' diverse backgrounds. TEE manufacturers such as Intel and AMD could incorporate this guidance into their developer documentation and whitepapers.
    \item \textbf{Adopt practices for reducing non-determinism.} Although the common sources of non-determinism are well known~\cite{DBLP:journals/cacm/DelignatLavaudFVCRCR24/CCFreproducibleBuildArticle}, many projects still rely on them (cf.~\cref{subsec:pitfalls}), and two interviewees reported substantial team effort spent tracking such issues down. We recommend established best practices, such as avoiding the \texttt{\_\_DATE\_\_} and \texttt{\_\_FILE\_\_} macros and using compiler options like GCC's \texttt{-ffile-prefix-map} for location-independent builds. Addressing these issues early can substantially improve reproducibility and avoid costly revisions later.
    \item \textbf{Document the build process and dependencies.} Our analysis shows that many open-source TEE projects lack sufficient documentation to reproduce enclave measurements, even when the required information is available to the developers. We recommend documenting the build process and its dependencies systematically, including exact versions and cryptographic hashes, and using standardized formats for build metadata. Debian's \texttt{.buildinfo} files~\cite{debian}, for example, record build environments, toolchains, and parameters, enabling verification across systems.
\end{enumerate}

\subsection{Organizational Managers and Stakeholders}

Beyond individual practices, our analysis highlights organizational barriers to reproducibility. We therefore propose the following recommendations for project managers and decision-making bodies:
\begin{enumerate}[leftmargin=*]
    \item \textbf{Prioritize reproducibility during development.}
    Several interviewees reported that reproducibility is not an organizational priority, and developers in large organizations often lack the authority to modify standardized CI/CD workflows.
    We therefore recommend treating reproducibility as a release requirement, supported by dedicated training, coding guidelines, and process audits. One participant's organization already enforces reproducibility as a prerequisite for deployment acceptance, a practice we argue should extend to all TEE-based applications.
    \item \textbf{Enforce code provenance and transparency.}
    Users currently must either trust developers to publish binaries built from the correct source or verify the correspondence between source and binary manually, which is error-prone.
    We recommend that organizations adopt code provenance services to improve transparency and accountability.
    Delignat-Lavaud \emph{et al.}~\cite{DBLP:journals/cacm/DelignatLavaudFVCRCR24/CCFreproducibleBuildArticle}, for example, propose a code transparency service that tracks TEE code provenance through a public, append-only ledger. Integrated into CI pipelines, such services can verify provenance automatically before release.
    \item \textbf{Standardize build environments and development workflows.}
    Half of the developers in our study reported challenges from inconsistent build environments, particularly in large, multi-team projects with external dependencies.
    Tools such as Docker~\cite{docker}, Nix~\cite{nix}, and Bazel~\cite{bazel} isolate the build process and ensure consistent toolchains. Docker encapsulates dependencies and configurations in portable container images, although versions must be pinned explicitly. Bazel sandboxes build actions and tracks input hashes, and its \emph{hermeticity} mode pins tools and dependencies. Nix is designed for reproducibility from the outset, enforcing exact dependency sources and hashes.
\end{enumerate}

\subsection{End Users and Customers}
Our findings suggest that the main obstacle to reproducibility is not technical difficulty but a lack of organizational incentive.

End users and customers can therefore drive adoption by requiring reproducibility explicitly in procurement specifications, requesting verifiable build information, and favoring products built with reproducible development practices. Such market pressure can elevate reproducibility from an optional feature to a core component of trust and security in TEE-based software.

\subsection{Special Interest Groups and Researchers}
Finally, special interest groups such as the IETF Supply Chain Integrity, Transparency, and Trust (SCITT) Working Group~\cite{scitt} can advance reproducibility in the TEE ecosystem by defining standards, tools, and best practices, and by raising awareness in developer and policy communities.
Researchers can complement these efforts by identifying reproducibility barriers systematically, developing improved verification mechanisms, and building open-source tools and reference implementations that make reproducibility practical and accessible across the ecosystem.

\section{Conclusion}

In this paper, we investigated a fundamental aspect of the security guarantees offered by TEEs: the reproducibility of TEE builds. We empirically analyzed 115 TEE deployments and conducted an interview study with 12 SGX developers from both industry and academia. Our analysis revealed that 91\% of the examined projects were not reproducible, indicating that reproducible builds remain the exception rather than the norm in today's TEE ecosystem.

Our findings indicate that reproducibility challenges are largely TEE-agnostic. Instead, they stem from a combination of technical and organizational factors, including non-deterministic build processes, inadequate documentation, limited control over build environments, and low prioritization.  
Interestingly, developers often perceive TEE reproducibility as inherently difficult. Yet, both prior research and our findings indicate that many sources of non-determinism in build processes are well-understood and can be addressed using straightforward technical solutions. This points to organizational and awareness-related factors as the primary obstacles to achieving reproducibility.

Building on these insights, we proposed targeted recommendations for TEE ecosystem stakeholders—developers, project managers, and standardization bodies—to raise awareness, establish reproducibility as a core quality and security requirement from project inception, and offer practical guidance for achieving it without disrupting existing workflows.

\begin{acks}
This work is partly funded by the Deutsche Forschungsgemeinschaft (DFG, German Research Foundation) under Germany’s Excellence Strategy - EXC 2092 CASA - 390781972, DFG project number 560392681, the European Union’s Horizon 2020 research and innovation program (REWIRE, Grant Agreement No. 101070627), and an ARC Discovery Project number DP210102670. 

Views and opinions expressed are, however, those of the authors only and do not necessarily reflect those of the European Union. Neither the European Union nor the granting authority can be held responsible for them.

Moreover, this paper was edited for grammar using Grammarly, DeepL, and ChatGPT.
\end{acks}

\bibliographystyle{ACM-Reference-Format}
\bibliography{references}

\appendix  

\section{Open Science}  

In compliance with the open-science policy, we make all research artifacts required to evaluate the contributions of this paper publicly available at~\cite{our_artifact}.

\paragraph{Artifacts}

The following artifacts are provided:
(i) a table containing the detailed results of our analysis of TEE projects from~\cite{awesome_sgx,azure,gcp};
(ii) detailed instructions on our reproducibility tests for the 27 applications listed in \Cref{tab:study}; and
(iii) scripts used to process the data and generate the tables in the paper.

\section{Ethical Considerations}

At the time of our study, our institution has no institutional review board (IRB) or ethics review board (ERB). However, we complied with strict institutional and \mbox{(inter-)}national data protection and privacy regulations. The research team included two experts from the Human-Centered Security department, who supervised the planning and ensured compliance with ethical standards for studies that involve humans. We used a consent form, which our data protection officer had previously verified. The consent form covered all information that would usually be required for IRB/ERB approval. Participants were informed that participation was voluntary and that they could withdraw from the study at any time without any consequences. Our contact information was provided in case participants had any questions regarding data protection or the survey. %
After identifying relevant projects, we used email addresses from GitHub to contact the experts and ask for participation in our study. We paid participants €50 for the 25-minute interview, which corresponds to €120 per hour. Before recording, all participants were reminded of the privacy policy and asked if they had any questions. All interviews were transcribed locally and then anonymized  to ensure that no sensitive information was published.

\section{Interview Guide}
\label{apx:interviewguide}

\noindent\textbf{\textit{1. Onboarding}}

\begin{itemize}
    \item Introduction of interviewers and the institution
    \item Motivation: Exploring the reproducibility of open-source SGX applications
    \item Introduction of the interviewee
    \item Clarification of any questions regarding the consent form and data protection
    \item Beginning audio recording (with consent)
\end{itemize}

\bigskip
\noindent\textbf{\textit{2. Narrative Prompt}}

\noindent We’ll begin with your general thoughts on reproducibility, especially in the context of your most recent SGX project.

\begin{enumerate}
    \item \textbf{Introduction Reproducibility}
    \begin{itemize}
        \item What does reproducibility mean to you when working with SGX?
        \item How did you first learn about reproducibility?
    \end{itemize}
    
    \item \textbf{Clarifying Terminology}  
    \begin{itemize}
        \item For this study, reproducibility refers to the ability to reproduce the enclave hash, linking source code to its binary output.
    \end{itemize}

    \item \textbf{Reproducibility in Your Current Project}
    \begin{itemize}
        \item Did you consider reproducibility in development?
        \item How important is reproducibility in this project? Why?
        \begin{itemize}
            \item Is it important to your users, community, or other stakeholders?
        \end{itemize}
        \item What steps have you taken to support reproducibility?
        \item Have you tested for reproducibility? If so, how?
        \item How much effort are you willing to invest (or have already invested) in ensuring reproducibility?
    \end{itemize}
\end{enumerate}

\bigskip
\noindent\textbf{\textit{3. Challenges in Reproducible Enclave Builds}}

\noindent Let’s now discuss obstacles you may have encountered in building reproducible TEE applications.

\begin{enumerate}
    \item \textbf{Project-Specific Challenges}
    \begin{itemize}
        \item What challenges have you faced in achieving reproducibility in your current project?
    \end{itemize}

    \item \textbf{Wider Context}
    \begin{itemize}
        \item We observed that reproducibility is often problematic across many open-source SGX projects.
    \end{itemize}

    \item \textbf{General and Cross-TEE Challenges}
    \begin{itemize}
        \item Have you experienced reproducibility issues in other projects? Which?
        \item Why do you think reproducibility is not widely addressed?
        \begin{itemize}
            \item Who could make reproducibility a priority?
        \end{itemize}
        \item Have you worked with other TEEs (e.g., TrustZone, TDX, SEV)?
        \begin{itemize}
            \item Have you encountered and addressed reproducibility challenges in those platforms?
            \item What are the key differences between SGX and other TEEs regarding reproducibility?
        \end{itemize}
    \end{itemize}
\end{enumerate}

\bigskip
\noindent\textbf{\textit{4. Improving Reproducibility}}

\noindent In this final section, we’d like to explore ideas and solutions to support more reproducible enclave builds.

\begin{itemize}
    \item How would you address the challenges you mentioned?
    \item One idea we’re exploring is using Docker containers to simplify reproducibility testing. Would this be feasible for your project? Do you think it would help others?
    \item Through which channels would you share reproducibility-related information (e.g., Intel’s website, GitHub, documentation)?
    \item How can the research community better support developers in achieving reproducibility?
\end{itemize}

\bigskip
\noindent\textbf{\textit{5. Offboarding}}

\begin{itemize}
    \item Do you have any final questions or topics you’d like to discuss?
    \item Concluding the session and stopping the recording
\end{itemize}

\section{Emails \& Codebook}
\label{apx:codebook}

\begin{strip}

    \centering
    \small
    \scalebox{0.9}{
    \begin{tcolorbox}[colback=gray!5!white,
                  colframe=gray!70!black,
                  title=Interview recruiting email,
                  boxrule=0.5pt,
                  arc=2pt,
                  left=3mm, right=3mm, top=2mm, bottom=2mm]

    \textbf{Subject:} Chat on reproducibility (SGX) - Academics need help\\[6pt]
    Dear \emph{Recepient},\\[6pt]
    
We are researchers from \emph{Institution, Country}, currently conducting a study on the reproducibility of open-source projects that use Intel SGX.\\

While exploring various projects on GitHub, we found that the majority of SGX-based projects were not easily reproducible---a challenge we are aiming to understand more deeply.\\

To that end, we are reaching out to project maintainers, developers, and contributors like yourself to hear your perspective on reproducibility---its relevance, challenges, and how it fits into your development process. Even if you are not familiar with reproducibility, or if it was not a specific focus for your projects, your input would still be extremely valuable to our study.\\

Would you be available for a brief interview ($\sim$25 minutes) at your convenience? Our goal is to support the open-source community in building more secure, reliable, and trustworthy software. In appreciation, we will provide you with a 50€ ($\sim$\$58) gift card from [xxxxx] after the interview, which you can redeem for vouchers covering a broad selection of services. \\

If you are interested, please fill out a short ($\sim$5 minutes) pre-survey and book a slot for the interview here: \emph{link-to-pre-survey}.\\

If you have any questions, feel free to reply to this email.\\

Best regards,\\
\emph{Author names}

    \end{tcolorbox}
    }
    \captionof{figure}{Email to SGX project developers and maintainers for interview regarding reproducibility challenges.}
    \label{fig:interview_email}

\end{strip}

\onecolumn
{
\scriptsize
\centering
\renewcommand{\arraystretch}{1.1}
\setlength{\tabcolsep}{0.6\tabcolsep}
\setlength{\defaultaddspace}{0.33\defaultaddspace} %
\rowcolors{2}{white}{gray!10}

\begin{xltabular}{\textwidth}{|lllX>{\raggedright\itshape\arraybackslash}p{6cm}l|}
\hiderowcolors
\caption{Codebook. (*) denotes a container for sub-codes and therefore is not used during coding. The themes map to the discussion in Section~\ref{dis:Challenges} and Section~\ref{sec:discussion_badge}: Organizational Barriers (OB), Known Issues (KI), and Security is a Badge (SB).}\label{tab:codebook-tee-reprod}\\
\hline
\endfirsthead

\hline

\multicolumn{3}{|l}{\textbf{Code}} & \textbf{Description} & \textbf{\textup{Example Quote}} & \textbf{\textup{Theme}}\\
\hline
\endhead

\showrowcolors

\hline
\endfoot
\hline
\endlastfoot

        \multicolumn{3}{|l}{\textbf{Code}} & \textbf{Description} & \textbf{\textup{Example Quote}} & \textbf{\textup{Theme}}\\
        \hline
        \multicolumn{3}{|l}{\textbf{TEE Technologies (*)}} & Subcategories are used for cross-code correlations to support sensemaking and the structuring of results. & --- & \\
            {} & \multicolumn{2}{l}{\textbf{Intel SGX}} & Coded if SGX or related technologies are explicitly mentioned. & \cquote{But yeah, SGX is part of that, but also AMD's TEEs as well.} & \\
            {} & \multicolumn{2}{l}{\textbf{Intel TDX}} & Coded if Intel TDX or related tech. are explicitly mentioned.  & \cquote{But when you go to SEV-SNP or on TDX, it's, it's more complex [...]} & \\
            {} & \multicolumn{2}{l}{\textbf{AMD SEV}} & Coded if AMD SEV or related tech. are explicitly mentioned.  & \cquote{I have a reasonably good estimate actually for SNP because this was, this is work that was mostly done by an intern recently. [...]} & \\
            {} & \multicolumn{2}{l}{\textbf{Arm TrustZone}} & Coded if Arm TrustZone or related tech. are explicitly mentioned. & \cquote{I mean, so the TrustZone, ArmCCA has attestation. [...]} & \\
            {} & \multicolumn{2}{l}{\textbf{Arm CCA}} & Coded if Arm CCA or related tech. are explicitly mentioned. & \cquote{ [...] it cannot do attestation. So CC can, ArmCCA. [...]} & \\
        \multicolumn{3}{|l}{\textbf{{Perceptions of Reproducibility} (*)}} & General attitudes, beliefs, and understandings of what reproducibility means in the participants' context. & --- & \\
            {} & \multicolumn{2}{l}{\textbf{General Relevance of Reproducibility}} & This code captures participants’ opinions, including statements that directly assess or describe the relevance of reproducibility, as well as those that imply its perceived importance indirectly. & \cquote{And so I think it's really important that you have the ability to do that.} & SB \\
            {} & \multicolumn{2}{l}{\textbf{Definition of the Concept}} & Statements in which participants describe how they define or understand reproducibility. & \cquote{The ability to create the same results from the same source codes [...]} & KI \\
        \multicolumn{3}{|l}{\textbf{{Prioritization of Reproducibility} (*)}} & Descriptions of how reproducibility is actually handled in day-to-day work, projects, or organizations. & --- & \\
            {} & \multicolumn{2}{l}{\textbf{Reproducibility is a priority}} & Situations where reproducibility is treated as an explicit goal or priority, including dedicated efforts or resources. & \cquote{It's like one of our major goals.} & SB \\
            {} & \multicolumn{2}{l}{\textbf{Reproducibility not prioritized}} & Situations where reproducibility is downplayed, neglected, or consciously traded off against other concerns. & \cquote{They know this is a problem, but they don't care about this. They don't want to invest manpower.} & SB \\
            {} & \multicolumn{2}{l}{\textbf{Relevance of Reproducibility (*)}} & Explanations of why reproducibility matters (or does not) in specific contexts, including stakeholders and use cases. & --- & \\
                {} & {} & \textbf{Parties Requesting Reproducibility} & Mentions of actors who demand or expect reproducibility & \cquote{Mostly customers.} & SB \\
                {} & {} & \parbox[t]{2.75cm}{\textbf{Parties Who Can Make \newline Reproducibility a Priority}} & Statements about people or roles who have the power to prioritize or enforce reproducibility. & \cquote{In the end, it's the customers, right.} & SB \\
                {} & {} & \parbox[t]{2.75cm}{\textbf{Use Cases Where \newline Reproducibility is Important}} & Concrete situations or scenarios in which reproducibility is particularly critical (e.g., incident response, certification, debugging). & \cquote{during remote attestation and all that [...]} & SB \\
        \multicolumn{3}{|l}{\textbf{{Challenges} (*)}} & Descriptions of obstacles and problems that hinder reproducibility. & --- & \\
            {} & \multicolumn{2}{l}{\parbox[t]{2.75cm}{\textbf{Challenges in Achieving \newline Reproducibility (*)}}} & Specific factors that make it difficult to obtain reproducible builds. & --- & \\
                {} & {} & \textbf{Timestamps} & Problems caused by time-dependent data (e.g., embedded timestamps) that lead to non-identical outputs. & \cquote{You also provide a build timestamp, which does change.} & KI \\
                {} & {} & \parbox[t]{2.75cm}{\textbf{(Absolute) paths}} & Issues due to absolute or environment-specific file paths affecting build or runtime outcomes. & \cquote{Paths is a really common issue. For example [...]} & KI \\
                {} & {} & \parbox[t]{2.75cm}{\textbf{Signatures are \newline non-deterministic}} & Challenges where cryptographic signatures or similar artifacts vary between builds and break reproducibility. & \cquote{[...] you have signatures that you have embedded in the binary that can also affect reproducibility.} & KI \\
                {} & {} & \textbf{Lack of awareness} & Descriptions of situations where stakeholders are unaware of reproducibility issues or their implications. & \cquote{So this is the first time I'm actually aware of this being an issue in production.} & SB \\
                {} & {} & \textbf{High complexity} & Difficulties stemming from complex systems, toolchains, or configurations that are hard to control or replicate. & \cquote{SGX development is notoriously complex.} & KI \\
                {} & {} & \parbox[t]{2.75cm}{\textbf{Hotfix release  before fix}} & Problems caused by rapid hotfixes or ad-hoc releases that bypass reproducible processes or documentation. & \cquote{[...] if you find a security issue and you immediately patch it in your TEE's and you immediately release the source code for reproducibility, you are potentially making things easier for attackers [...]} & OB \\
                {} & {} & \parbox[t]{2.75cm}{\textbf{Debugging Non-\newline Reproducibility is hard}} & Statements emphasizing that diagnosing the causes of non-reproducible behavior is time-consuming or technically challenging. & \cquote{the hardest thing I think for reproducibility is the debugability of it} & KI \\
                {} & {} & \parbox[t]{2.75cm}{\textbf{Limited Control of \newline the Environment}} & Difficulties arising from partial control over hardware, OS, toolchain, or external services that affect reproducibility. & \cquote{in my experience a lot of moving parts, system or program which is complex enough has so many moving parts, it's pretty hard to count them all in [...]} & OB \\
                {} & {} & \parbox[t]{2.75cm}{\textbf{Challenges are often \newline not TEE specific}} & Observations that many obstacles stem from general software-engineering issues rather than TEEs themselves. & \cquote{They're not really specific to SGX or to confidential computing in general.} & OB \\
                {} & {} & \parbox[t]{2.75cm}{\textbf{Transferrability to \newline other TEEs}} & Concerns about whether reproducibility-related solutions or problems transfer across different TEE platforms. & \cquote{I think basically it's the same problem} & OB \\
            
        \multicolumn{3}{|l}{\textbf{Dealing with challenges (*)}} & Ways in which participants respond to or try to manage the above challenges. & --- & \\
            {} & \multicolumn{2}{l}{\textbf{{Practices to Achieve Reproducibility}}} & Concrete methods, workflows, or tools used to improve or enforce reproducibility in practice. & \cquote{build the enclave inside the same Docker container environment on both the Attester and Relying Party sides} & KI \\
            {} & \multicolumn{2}{l}{\parbox[t]{2.75cm}{\textbf{{Suggestions for Improvement} \newline {of Reproducibility} (*)}}} & Proposed ideas or wishes for improving reproducibility. & --- & \\
                {} & {} & \textbf{Technical Improvements} & Suggestions for technical changes (e.g., tooling, build systems, TEE features) that would facilitate reproducibility. & \cquote{pinning the Docker, like pinning this strict version, I mean, that's what SBOM is for} & KI \\
                {} & {} & \parbox[t]{2.75cm}{\textbf{General Desirable \newline Improvements}} & Broader improvement suggestions, e.g., better processes, documentation, or organizational practices around reproducibility. & \cquote{we should increase the awareness. [...]} & OB, SB \\
                {} & {} & \textbf{What Research Can Do} & Expectations or wishes regarding contributions from academic or industrial research to support reproducibility. & \cquote{Having researchers directly join projects as maintainers could improve reproducibility quality [...]} & OB, SB \\
                {} & {} & \parbox[t]{2.75cm}{\textbf{Channels for Spreading Reproducibility-related Information}} & Mentions of how knowledge about reproducibility should be disseminated (e.g., documentation, talks, communities, standards). & \cquote{For more general concerns, I often use technical blogging platforms [...] or my company’s engineering blog.} & SB \\
\end{xltabular}

}

\end{document}